\documentclass[sigconf]{acmart}

\usepackage{booktabs} 
\usepackage{balance}
\usepackage{subfigure}

\usepackage{listings}
\usepackage{xcolor}

\colorlet{punct}{red!60!black}
\definecolor{background}{HTML}{EEEEEE}
\definecolor{delim}{RGB}{20,105,176}
\colorlet{numb}{magenta!60!black}
\lstdefinelanguage{json}{
    basicstyle=\normalfont\ttfamily,
    numbers=left,
    numberstyle=\tiny,
    stepnumber=1,
    numbersep=8pt,
    showstringspaces=false,
    breaklines=true,
    frame=lines,
    backgroundcolor=\color{background},
    literate=
     *{0}{{{\color{numb}0}}}{1}
      {1}{{{\color{numb}1}}}{1}
      {2}{{{\color{numb}2}}}{1}
      {3}{{{\color{numb}3}}}{1}
      {4}{{{\color{numb}4}}}{1}
      {5}{{{\color{numb}5}}}{1}
      {6}{{{\color{numb}6}}}{1}
      {7}{{{\color{numb}7}}}{1}
      {8}{{{\color{numb}8}}}{1}
      {9}{{{\color{numb}9}}}{1}
      {:}{{{\color{punct}{:}}}}{1}
      {,}{{{\color{punct}{,}}}}{1}
      {\{}{{{\color{delim}{\{}}}}{1}
      {\}}{{{\color{delim}{\}}}}}{1}
      {[}{{{\color{delim}{[}}}}{1}
      {]}{{{\color{delim}{]}}}}{1},
}

\begin{document}
	
	
\title{MQTT+: Enhanced Syntax and Broker Functionalities for Data \\ Filtering, Processing and Aggregation}

\author{Riccardo Giambona}
\affiliation{%
  \institution{DEIB - Politecnico di Milano}
  \streetaddress{Piazza Leonardo da Vinci 32}
  \city{Milano}
  \state{Italy}
  \postcode{20133}
}
\email{riccardo.giambona@mail.polimi.it}

\author{Alessandro E. C. Redondi}
\affiliation{%
  \institution{DEIB - Politecnico di Milano}
  \streetaddress{Piazza Leonardo da Vinci 32}
  \city{Milano}
  \state{Italy}
  \postcode{20133}
}
\email{alessandroenrico.redondi@polimi.it}

\author{Matteo Cesana}
\affiliation{%
  \institution{DEIB - Politecnico di Milano}
  \streetaddress{Piazza Leonardo da Vinci 32}
  \city{Milano}
  \state{Italy}
  \postcode{20133}
}
\email{matteo.cesana@polimi.it}

%
%
%
%

\begin{abstract}
In the last few years, the Message Queueing Telemetry Transport (MQTT) publish/subscribe protocol emerged as the de facto standard communication protocol for IoT, M2M and wireless sensor networks applications. Such popularity is mainly due to the extreme simplicity of the protocol at the client side, appropriate for low-cost and resource-constrained edge devices. Other nice features include a very low protocol overhead, ideal for limited bandwidth scenarios, the support of different Quality of Services (QoS) and many others. However, when an edge device is interested in performing processing operations over the data published by multiple clients, the use of  MQTT may result in high network bandwidth usage and high energy consumption for the end devices, which is unacceptable in resource constrained scenarios. To overcome these issues, we propose in this paper MQTT+, which provides an enhanced protocol syntax and enrich the pub/sub broker with data filtering, processing and aggregation functionalities. MQTT+ is implemented starting from an open source MQTT broker and evaluated in different application scenarios.
\end{abstract}

%
%



\maketitle
\section{Introduction}\label{sec:introduction}
The Internet of Things (IoT) is day by day becoming a reality. Tiny and cheap devices equipped with sensors and wireless communication capabilities are being used more and more frequently in several application scenarios, such as wireless sensor networks, environmental monitoring, e-health, etc. Regardless of the specific scenario, all IoT applications are characterised by common requirements: sensor nodes operate with low-bandwidth wireless transceivers to transmit/receive data to/from a common data concentrator (sink node) or other IoT nodes. Such data may be then processed, according to the application's needs, either at the sink node or onboard other sensor nodes using low-power and energy-efficient micro-controllers. Such a resource-constrained environment stimulated in the last few years a vast body of research to design and optimise existing protocols at all layers of the communication stack. For what concerns the application layer, several efforts have been performed: protocols such as the Message Queue Telemetry Protocol (MQTT)\cite{banks2014mqtt}, the Constrained Application Protocol (COAP)\cite{bormann2012coap} and the Extensible Messaging and Presence Protocol (XMPP)\cite{saint2009xmpp} are the results of such efforts.

Among the existing solutions, MQTT is certainly the one that has received the greatest attention in the last few years, practically becoming the standard \textit{de-facto} in M2M and IoT applications. As a matter of fact, MQTT is becoming the most popular protocol to connect resource constrained devices to the major cloud platforms (e.g., Amazon AWS, Microsoft Azure, IBM Watson), which all expose their services through MQTT. The reasons of such popularity derive from MQTT's incredible simplicity client-side, which nicely fits in resource-constrained applications, yet supporting reliability and several degrees of quality of service (QoS). 
MQTT is based on the publish/subscribe pattern, and all communications between nodes are made available by a broker. The broker accepts messages published by devices and forwards them to clients subscribed to those messages, ultimately controlling all aspects of communication between devices. 

There is however a set of common IoT and M2M applications scenarios where the use of MQTT causes an inefficient use of the available network and computing resources. Those are all cases where data consumers (subscribers) are interested in only a subset of the data produced (published) by sensor devices, while the broker still forwards the entire data available. Examples include clients interested in receiving data only if it respects some condition, clients interested in certain aggregation functions (e.g., cumulative sum, average) over a set of data published, or clients interested in the result of some processing task over such data, rather than the data itself. In all these case, two main drawbacks can be identified: (i), the data forwarded by the broker may potentially be discarded by subscribers, wasting network resources and (ii) subscribers need to perform additional processing operations, consequently decreasing their available computational and energy resources. 

To mitigate those issues, we propose in this paper an advanced MQTT broker (MQTT+) able to deal with such situations. MQTT+ allows a client to subscribe to advanced functionalities on the data published, including rule-based data filtering, spatial and temporal data aggregation and data processing. All functionalities are provided reusing as much as possible the original MQTT protocol logical and syntactical rules and minimally modifying the client-side procedures. As a proof of concept, MQTT+ is implemented starting from a publicly available MQTT broker and evaluated in different application scenarios.

The remainder of this work is structured as follows: Section \ref{sec:background} briefly discuss the MQTT protocol, highlighting its main features. Section \ref{sec:system_implementation} introduces MQTT+ and the proposed enhancements, while Section \ref{sec:system_evaluation} evaluates it in different scenarios. Section \ref{sec:related_work} summarises related works dealing with MQTT enhancements and in general in the area of publish/subscribe middlewares. Finally, Section \ref{sec:conclusion} concludes the paper and discusses future work directions.

\begin{table*}[t]
\centering
\caption{MQTT+ rule based operators}
\label{tab:rule_based_operators}
\begin{tabular}{|l|c|l|}
\hline
\multicolumn{1}{|c|}{\textbf{Rule-based subscription}} & \textbf{Value Type} & \multicolumn{1}{c|}{\textbf{Description}}                                                \\ \hline
\verb|$EQ;value/topic|          & numeric,string             & Forwards data to subscriber if data published on \verb|topic| is equal to \verb|value|             \\ \hline
\verb|$NEQ;value/topic|          & numeric,string             & Forwards data to subscriber if data published on \verb|topic| is different from \verb|value|             \\ \hline
\verb|$GT;value/topic|          & numeric             & Forwards data to subscriber if data published on \verb|topic| is greater than \verb|value|             \\ \hline
\verb|$GTE;value/topic|         & numeric             & Forwards data to subscriber if data published on \verb|topic| is greater than or equal to \verb|value| \\ \hline
\verb|$LT;value/topic|          & numeric             & Forwards data to subscriber if data published on \verb|topic| is less than \verb|value|                \\ \hline
\verb|$LTE;value/topic|         & numeric             & Forwards data to subscriber if data published on \verb|topic| is less than or equal to \verb|value|    \\ \hline
\verb|$CONTAINS;text/topic|     & string              & Forwards data to subscriber if data published on \verb|topic| contains \verb|value|                    \\ \hline
\end{tabular}
\end{table*}

\section{MQTT protocol overview}\label{sec:background}
The Message Queuing Telemetry Transport is a lightweight publish/subscribe protocol whose design principles are to minimise both the end-devices requirements and the utilised network resources, still ensuring reliability and some degree of quality of service.

MQTT follows a traditional publish/subscribe pattern in which a \textit{client} device publishes information relative to a particular \textit{topic}, i.e., a multilevel string describing the data being published (e.g. \verb|kitchen/temp|). Other clients interested in such information subscribe to that topic. Information forwarding from the publishers to the subscribers is made possible by a \textit{broker}, which is the core part of the system and is in charge of receiving data from the publishers and forwarding it to the subscribers. Such designs allow to decouple the publishing and subscribing processes: clients interested in a particular topic do not need to know who the publishers are, neither have they to be synchronised to the publishing operations. 

Before being able to publish data or subscribe to any topic, each client needs to connect to a broker. Such connection is based on TCP/IP and implemented through a simple message exchange between the client and the broker. During this process, a client communicates several information to the broker such as its client identifier, the connection keep alive time interval and other optional parameters (authentication, last will topic and message, etc). 

After the connection a client may directly start publishing data or subscribing to a certain topic using specific MQTT messages with minimal transport overhead (the fixed-length header is just 2 bytes). For both operations, clients have the possibility of choosing a Quality of Service (QoS) value, which impact on the way the broker handles the messages from/to the clients. Three QoS levels are defined: (i) at most once (fire-and-forget), which relies on the underlying TCP connection; (ii) at least once, where the sender will retransmit a message until an ACK is received and (iii) exactly once, where it is guaranteed that a message transmitted is received only once by the counterpart. Optionally, a client may publish \textit{retained} messages, by setting the retain flag to true during publication. Such messages will be stored internally by the broker and forwarded to any client subscribing to that message topic immediately after subscription. 

For what concerns the syntax of topic strings, the latest MQTT standard specifications\footnote{http://docs.oasis-open.org/mqtt/mqtt/v3.1.1/os/mqtt-v3.1.1-os.pdf} allow to use single or multilevel case-sensitive topics, where each level is separated by a forward slash. Each topic must have at least one character to be valid and a broker accepts each valid topic without and prior initialisation. A client may subscribe to a specific topic by using the exact topic string, or subscribe to multiple topics at once by using a single-level (+) or multi-level (\#) wildcard. The broker will then forward to the client all messages whose topic matches the subscription topic, including the wildcard. As an example, a client subscribing to \verb|kitchen/#| will receive messages published on both the \verb|kitchen/temp| and \verb|kitchen/hum| topics, while a client subscribing to \verb|+/hum| will receive messages published on both the \verb|bathroom/hum| and \verb|kitchen/hum| topics. Additionally, the standard specifies that topics beginning with the character \verb|$| are reserved for special uses and cannot be utilised by client applications for publishing data. In particular, \verb|$SYS/| has been widely adopted by most of the publicly available MQTT broker implementations as a prefix to topics that contain broker-specific information. As an example, a client may subscribe to \verb|$SYS/broker/clients/connected| to receive the number of currently connected clients.


\section{MQTT+ enhanced functions}\label{sec:system_implementation}
As explained in Section \ref{sec:background}, one important feature of the MQTT protocol is to be notably lightweight client-side. This is mainly due to the presence of the broker, which is responsible of the most intensive operations of the protocol. The proposal of this work is to take another step in this direction, by adding several functionalities to the broker in order to further decrease the computational complexity of the clients and the overall network resources utilisation. Several additional functionalities are provided by MQTT+: rule-based subscriptions, temporal/spatial data aggregation and intensive data processing. All these functions are oriented at decreasing the computational load on clients and on the network segment from the broker to the subscribers, at the cost of a slight increase in the complexity of the broker implementation. In order to leverage such functions, an enhanced syntax is introduced. The new syntax is nicely integrated with the original MQTT syntactic rules by making use of leading \verb|$| characters followed by several specific keywords and requires optional modifications of negligible impact on clients. Indeed, MQTT+ is completely backward compatible with standard MQTT devices.

\subsection{MQTT+ rule based subscription}
In many cases, a client is interested in a topic only if the data published on it respect some condition. As an example, consider an automatic alarm device which needs to fire only if the measurement provided by a temperature sensor (say \verb|sens123|) is greater than a certain threshold. Of course, the alarm device may subscribe to the \verb|sens123/temp| topic and than process internally the received data to decide when to react. However, the same behaviour can be more efficiently achieved if the broker knows at which condition a message should be forwarded and operates accordingly. In this case, the broker acts as a data filter, avoiding to forward unnecessary data and thus saving bandwidth. MQTT+ allows a client to perform a rule-based subscription using ad-hoc operators: referring to the previous example, the alarm device may subscribe to the \verb|$GT;value/sens123/temp| topic and receive only messages published on the \verb|sens123/temp| topic whose payload contains a value greater than \verb|value|. Similarly, operators for greater than or equal (\verb|$GTE;value|), less than (\verb|$LT;value|) and less than or equal (\verb|$LTE;value|) are defined. All these operators require the data published on the topic subject of the subscription to be numeric: an MQTT+ broker will accept any numeric rule-based subscription but will ignore it if the value published is not numeric. At the same time, we observe that MQTT messages are often used to carry non-numeric data such as JSON or XML documents. MQTT+ allows rule-based operation also on such non-numeric payloads: the \verb|$CONTAINS;text| operator is defined, which searches in the payload of the published message the string \verb|text|. The broker will forward such a message only if the string is found. Finally, the equality (\verb|$EQ;value|) and inequality (\verb|$NEQ;value|) operators are defined for both numeric values and string. MQTT+ rule-based operators are summarised in Table \ref{tab:rule_based_operators}.

\begin{table*}[t]
\centering
\caption{MQTT+ broker internal memory structure}
\label{tab:buffer}
\begin{tabular}{|c|c|c|c|c|c|c|c|c|c|c|c|c|c|}
\cline{1-8} \cline{10-14}
Topic               & Last Value     & TTL                 & $N_{T_{1440}}$ & $S_{T_{1440}}$ & $A_{T_{1440}}$ & $L_{T_{1440}}$ & $U_{T_{1440}}$ &          & $N_{T_{15}}$ & $S_{T_{15}}$ & $A_{T_{15}}$ & $L_{T_{15}}$ & $U_{T_{15}}$ \\ \cline{1-8} \cline{10-14} 
\verb|/sens1/temp|   & 22.5           & 2018-05-24T15:36:25 & 62             & 1063           & 17.14          & 12.3           & 24.1           &          & 2            & 45.3         & 22.65        & 22.5         & 22.8         \\ \cline{1-8} \cline{10-14} 
\verb|/sens2/temp|   & 13.8           & 2018-05-24T15:33:42 & 38             & 386            & 10.15              & 6.2              & 11.8              & $\cdots$ & 3            & 34.2            & 11.4           & 11.2            & 11.6            \\ \cline{1-8} \cline{10-14} 
\verb|/sens1/hum|    & 89.1           & 2018-05-24T15:36:26 & 61             & 4758           & 71              & 69              & 89.1              &          & 2            & 178          & 89            & 88.9            & 89.1          \\ \cline{1-8} \cline{10-14} 
\verb|/sens1/status| & \{status: ok\} & 2018-05-24T15:36:27 & -              & -              & -              & -              & -              &          & -            & -            & -            & -            & -            \\ \cline{1-8} \cline{10-14} 
\verb|$CNTPPL/image1| & 12 & 2018-05-24T15:32:12 & 10              & 103              & 10.3              & 8              & 12              &          & 1            & 12           & 12            & 12           & 12          \\ \cline{1-8} \cline{10-14} 
\end{tabular}
\end{table*}

\subsection{MQTT+ data TTL}
Upon data publication from a client on a topic, a traditional MQTT broker checks the list of subscribers to the topic and forwards the data to them. After a successful forwarding, according to the QoS set during publication, the original message is deleted, unless it was published with the retain flag set. In that case, the message is kept in memory in order to be forwarded to any new subscriber to that topic.
In MQTT+, we propose to slightly modify this paradigm and include in each published message a data Time To Live (TTL) information, a timestamp value (e.g., dd/mm/yyyy HH:MM:SS) which explicitly informs the broker of how long the data published should be considered valid, and therefore stored in the broker internal memory.
Operatively, two options are available for communicating the TTL field to the broker:
\begin{enumerate}
	\item Explicit TTL: the TTL field may be inserted in the variable part of the PUBLISH message header, which already carries important information such as the topic name, the QoS value and the message ID. The addition of a new field in the variable part of the header causes no issues since the length of the variable header is carried in the fixed part of the header.
	\item Implicit TTL: in case no explicit TTL is provided during publication, a default TTL (e.g. 1 hour from the publication timestamp) may be automatically assigned by the broker.
\end{enumerate}
As we shall see later, the TTL field is crucial for allowing correct data aggregation functionalities at the broker.

\subsection{MQTT+ temporal data aggregation}\label{sec:temporal_aggregation}
In many applications, a device is interested in obtaining data at a much lower frequency compared to the publication rate. As an example, a device for optimising the residential energy consumption may be interested only in the daily current consumption of certain household appliances, rather than obtaining a fine-grained time series from each one of them. Again, the broker could be exploited to provide temporal aggregation functionalities. For each topic, the MQTT+ broker not only stores the last published data with its TTL, but also keeps in memory $K$ tuples of the form ($N_{T_k}$, $S_{T_k}$,$A_{T_k}$,$U_{T_k}$,$L_{T_k}$), as illustrated in Table \ref{tab:buffer}. Each tuple stores, for the last ${T_k}$ minutes, the number of publish events $N_{T_k}$, the cumulative sum of the data value published $S_{T_k}$, the average value $A_T = S_{T_k}/N_{T_k}$ and the maximum and minimum value received, $U_{T_k}, L_{T_k}$. Although the intervals ${T_k}$ can be chosen arbitrarily, a reasonable choice could be to have $K = 3$ and the corresponding $k = \{1440,60,15\}$. In that case, the broker would store the daily, hourly and quarter-hourly statistics for each numeric topic\footnote{In case a non-numeric data is published on a topic, it is not considered valid for aggregation and only the last published value is stored}. A single timer expiring every $D$ minutes, where $D$ is the lowest common denominator among the ${T_k}$ periods, is needed to periodically reset such fields (e.g., the $T_{15}$ tuple is reset every time the timer fires while the $T_{60}$ tuple every four times). Also, note that the memory complexity of such structure is linear in the number of published topics.

With the proposed data structure, several types of temporal data aggregation subscriptions may be enabled. A client interested in receiving temporally aggregated data for the \verb|topic| topic may subscribe to \verb|$<TIME><OP>/topic|, where according to the proposed time granularities, \verb|TIME| = \verb|{DAILY,HOURLY,QUARTERHOURLY}| and \verb|OP| = \verb|{COUNT,SUM,AVG,MIN,MAX}| maps to the defined tuples. As an example, a client interested in subscribing to the daily average of the current consumption of a certain household, published e.g.. on \verb|sens123/currcons|, can inform the MQTT+ broker of such intention by subscribing to \verb|$DAILYAVG/sens123/currcons|. The broker will react to such a subscription by publishing the value $A_{T_{1440}}$ on the  \verb|$DAILYAVG/sens123/currcons| topic, each time the corresponding timer fires (in this case every 96$D$ minutes, i.e., once per day).




\subsection{MQTT+ spatial data aggregation}
Besides temporal aggregation, a client may be interested in spatially aggregating several topics at once. As an example, the device for optimising residential energy consumption mentioned in the previous section may be interested in obtaining the sum of the energy consumption of the single appliances directly from the broker, rather than computing it onboard. MQTT+ allows a client to subscribe to several aggregating functions over multiple topics. The topics to be aggregated can be specified either using standard MQTT wildcards or with an explicit syntax.
\subsubsection{Spatial aggregation with wildcards}
In case a client is interested in aggregating all data matching a certain topic name, it may subscribe to \verb|$<OP>/topic/|, where \verb|OP| can assume the same values defined for temporal aggregation and \verb|/topic/| contains one or more wildcards, according to the original MQTT rules. Only messages published with numeric data will be considered, and the aggregation function will be executed each time a new data is published on any topic matching the subscription. As an example, consider two temperature sensors publishing on the \verb|room1/sens1/temp| and \verb|room1/sens2/temp| topics. A client interested in computing the average of the two values may subscribe to \verb|$AVG/room1/+/temp|. Every time a new data is published by any sensor, the broker will perform the average of the last published values of all topics matching \verb|room1/+/temp|. Note that each client publishes a message asynchronously and independently of each other. To prevent that data with referring to different time instants is aggregated, the MQTT+ broker considers only those entries whose TTL is valid. Note also that, as long as data is numeric and has a valid TTL, the broker will compute the aggregation function as requested, without any further check on the topics being aggregated. As an example, considering the published topics \verb|room1/temp| and \verb|room1/hum|, the subscription to \verb|$AVG/room1/#| will produce a valid value, although meaningless. The correct use of the aggregation functions is therefore left to the final user.

An issue that occurs when subscribing to a spatial aggregation topic with wildcards is the choice of which topic to use for publication. In standard MQTT, a subscription to a topic with wildcards is equivalent to subscribing to all matching topics and the broker will forward data to the subscriber on each individual topic. In case of an aggregated subscription, one single topic must be used for publishing the aggregation results. One cannot use the same topic used for subscription, as MQTT rules do not allow to use a wildcard in a publication topic. Therefore, we propose two different possibilities for choosing such a topic. The two options differ in how the wildcard is replaced:
\begin{enumerate}
	\item \textit{Single keyword replacement (SKR):} wildcards may be replaced with the unique keyword \verb|$AGGREGATE| during the forwarding process to the broker. Referring to the previous example, subscribing to \verb|$AVG/room1/+/temp| will trigger the broker to reply on  \verb|$AVG/room1/$AGGREGATE/temp|.
	\item \textit{Replacement with participating topics (RPT):} in case of single keyword replacement, the subscribers is unable to understand which topics participated in the aggregation. An alternative possibility is for the broker to explicit insert such participating topics during the publish process. In this case, wildcards are replaced with the string \verb|t1;t2;...;tn|, obtained concatenating all the topics participating to the aggregation. Referring again to the previous example, the broker will reply on \verb|$AVG/room1/sens1;sens2/temp| if both values have valid TTL. The main drawback of this approach is that the length (in bytes) of the topic forwarded by the broker increases with the number of topics aggregated, therefore decreasing the aggregation efficiency.
\end{enumerate}
The choice of which option to use is left to the client during subscription and encoded into the subscribe messages payload.

\subsubsection{Explicit spatial aggregation}
MQTT+ also supports a different spatial aggregation operation, where the topics to be aggregated are explicitly communicated to the broker during subscription. In this case, a client uses the concatenation of all topics to be aggregated (e.g., \verb|t1;t2;...;tn|) during subscription. Referring to the previous example, subscribing to \verb|$AVG/room1/sens1;sens2/temp| will trigger the broker to average data coming from the two sensors. Also in this case, only the messages with valid TTL will be considered for aggregation and the broker will reply indicating only the topics corresponding to the considered ones (similar to the replacement with participating topics case in the case with wildcards).

\subsection{MQTT+ data processing}
One of the main features of MQTT is that practically any type of data can be transferred with publish and subscribe messages. With a maximum payload size of 256 MB, MQTT paves the way to advanced applications in which more complex data, rather than just numbers, are transmitted. An interesting case study which particularly fits in this scenario is the one of wireless surveillance cameras, which are nowadays more and more used. Consider one or more cameras that take images at specific time interval and transmit them over MQTT to a broker. We focus on the case where subscribers to such image topics are interested in the content of such images, rather than in the pixel-domain based representation of the images. As an example, a subscriber may be interested in counting how many people are present in an image, if a certain person is there or what kind of objects are present. All these operations require image analysis algorithms to be run on the subscriber devices after the broker has forwarded the images from the cameras. This has two important drawbacks: (i) a huge amount of bandwidth is used for transmitting the raw images to the subscribers even though such subscribers are only interested in their semantic content and (ii) the image analysis on the subscriber devices is in general computationally-eager and thus should be optimised.
The proposed enhanced MQTT+ allows to overcome such drawbacks by enabling data processing directly at the broker. We focus here only on image processing, although the framework can be extended to any other type of data and processing operations (video and data compression, signal processing, etc.). In particular, we focus on the scenario in which one  camera is available (e.g., publishing on \verb|/room1/image|), and subscribers are interested in counting how many people are present in the published images. To do this, the broker allows to subscribe to \verb|$CNTPPL/room1/image|. When an image is published on that topic, the broker runs a human detection algorithm and returns the number of estimated people to the subscriber. Note that the data processing functions depend only on what image analysis algorithms the broker is able to run. In principle, a broker may be even connected to a cloud-based web service enabling such processing functions (e.g., Amazon AWS Rekognition\footnote{https://aws.amazon.com/rekognition} or any other service).
In any case, a client willing to connect to a broker should know which processing functions are available. We propose to use one of the MQTT system topics (the one starting with \verb|$SYS|, e.g., \verb|$SYS/capabilities|) so that the broker can advertise its available processing functions. Upon subscription on such a topic, the MQTT+ broker replies with a JSON containing all available functions and a corresponding description (see Figure \ref{fig:json}).

\begin{figure}
	\centering
\begin{lstlisting}[language=json,firstnumber=1]
[{ "keyword": "$CNTPPL",
   "desc": "counts people in an image",	
   "returns": "value"
  },
  {"keyword": "$CNTMALE",		
   "desc": "counts males in an image",
   "returns": "value"	
  },	
  {		
   "keyword": "$CNTFEMALE",		
   "desc": "counts females in an image",		
   "returns": "value"	
  },	
  {
   "keyword": "$RECOGNIZE",		
   "desc": "recognizes objects in an image",	
   "returns": "json"	
}]
\end{lstlisting}
\caption{Example of MQTT+ broker reply to a subscription on the capabilities topic}
\label{fig:json}
\end{figure}

\subsection{MQTT+ composite subscriptions}
One of the strengths of MQTT+ is the capability of allowing composite subscriptions, by properly chaining the operators introduced so far thus enabling even more advanced functions. In particular MQTT+ allows the following composite subscriptions:
\begin{enumerate}
	\item \textit{Spatio-temporal aggregations:} any temporal aggregation can be also aggregated spatially. As an example, when subscribing to \verb|$SUM$DAILYAVG/+/temp|, the MQTT+ broker performs the following operations when the daily timer expires: (i) it identifies all matching topics after the operators (e.g., the ones matching +/temp); (ii) it fetches the daily average of such topics from the internal buffer and (iii) it publishes the aggregate using the spatial aggregator (sum in this case) and using either single keyword replacement or replacement with participating topics (according to the subscriber's choice). Note that a spatio-temporal aggregation to a single topic is equivalent to a temporal aggregation over that topic (\verb|$SUM$DAILYAVG/room1/temp| and \verb|$DAILYAVG/room1/temp| produce the same effect). Also, subscriptions in which the temporal aggregator appears before the spatial aggregator (e.g., \verb|$DAILYAVG$SUM/+/temp|) are not permitted, since they would produce meaningless results. In this case, according to the original MQTT specifications, the broker returns a subscription acknowledgement (SUBACK) message reporting a failure.
	\item \textit{Spatio-temporal aggregation of processed data:} similarly, any data produced by a processing function may be aggregated spatially, temporally or spatio-temporally. As an example, subscribing to \verb|$SUM$DAILYAVG$CNTPPL/+/image| has the following effect: (i) all images published on the matching topics are processed by the broker to extract the number of people present and stored as standard entry in the buffer, so that the temporal statistics can be updated (see last row of \ref{tab:buffer}); (ii) when the daily timer expires, the daily averages are aggregated using the \verb|$SUM| operators and forwarded to the subscribers.
	\item \textit{Rule-based spatio-temporal aggregation} finally, any type of aggregation (spatial,temporal or spatio-temporal) may be subject to rules. Such rules may appear either before or after the aggregation operator, thus working on the input or output of the aggregation functions. The following examples are valid subscriptions: (i) \verb|$GT;value$SUM$DAILYAVG/+/temp|, which forwards to the subscribers the sum of the average of all temperature sensors only if it is greater than \verb|value|; (ii) \verb|$SUM$GT;value$DAILYAVG/+/temp|, which aggregates daily averages only if they are greater than \verb|value|. Subscriptions in which the rule-based operator comes after the temporal aggregator (e.g., \verb|$SUM$DAILYAVG$GT;value/+/temp|) are not permitted and result in a SUBACK message reporting a failure.
\end{enumerate}

\section{Implementation and Experiments}\label{sec:system_evaluation}
We implemented the proposed MQTT+ broker syntax and functionalities starting from the HiveMQ\footnote{https://www.hivemq.com/} 3.4 broker implementation, which offers a free and open source Java SDK to modify the broker functionalities via plugins. The plugin SDK provides to the developer several \textit{callbacks}, which can be linked to user-defined logics. This allows to modify the general system behaviour depending on specific events. As an example, the \textit{OnPublishReceivedCallback} is executed whenever an MQTT PUBLISH message arrives at the broker. The MQTT+ implementation of such a callback is responsible to add the published topic to the internal buffer shown in Table \ref{tab:buffer} or, if already present, to update the topic statistics. Similarly, the \textit{OnSubscribeCallback} is called whenever a subscription is performed by a client connected to the broker. The implementation of such callback is therefore responsible of understanding and validating the advanced syntax used in the subscription, eventually managing it in a proper way.

\subsection{Simulation scenarios}
In order to test the correct functionalities of the system, a simulation environment is created. The framework allows to create several MQTT Clients based on the Eclipse Paho project\footnote{http://www.eclipse.org/paho/}, acting either as data publishers (sensors) or subscribers. Upon start up, (i) the MQTT+ broker is booted up, (ii )$m$ sensors and $n$ subscribers are created and (iii)  the publishing process is started according to specific time parameters. Each sensor can publish a specific type of data on an individual topic, with a  deterministic rate of $lambda$ messages/second and the simulation runs for $T$ seconds. During this time window, the simulator monitors continuously three main performance figures:
\begin{enumerate}
	\item \textit{Total downlink traffic:} the testing environment uses \textit{tshark} (the command line tool of the popular Wireshark packet analyser software) to monitor continuously the traffic outgoing the broker on the TCP port 1883, used by MQTT. Both the broker and the clients are executed on the same physical machine (an Intel i7-6700@3,4 GHz with 8 GB of RAM, running Windows 10): to simulate realistic conditions of operation, all TCP traffic outgoing port 1833 is first rerouted to a Wi-Fi access point (through manipulation of the routing tables).
	\item \textit{Normalised CPU Load:} the Windows PowerShell \textit{GetProcess} tool (similar to the Unix \textit{top} tool) is used to keep track of the CPU load on the broker. Let $C_{m,n}$ be the average CPU load of a broker when $m$ clients publish on individual topics and $n$ clients are subscribed to such topics. We take $C_{1,1}$ as a reference load value and define the Normalised CPU Load as:
	\begin{equation}
		C_{m,n} = \frac{C_{m,n}}{C_{1,1}}
	\end{equation}
	\item \textit{Average RAM Consumption:} the framework also keeps track of the average RAM memory usage (in MB) of the broker during its different working phases.
\end{enumerate}
The first performance figure is crucial to understand the benefits of MQTT+ on the network resources, while the latter two performance measures are utilised to analyse the impact that the advanced functions of MQTT+ have on the complexity of the broker. This is important, considering that in many IoT and M2M application the broker may be implemented on low cost and low-power hardware platforms (e.g., Raspberry PI).

Two different application scenarios are considered, where sensors publish different types of data and subscribers are interested in different functions over such data: (i) spatio-temporal aggregation of scalar measurements and (ii) processing of image data.

\begin{figure*}[t!]
	\centering
	\subfigure[]{\includegraphics[width=0.66\columnwidth]{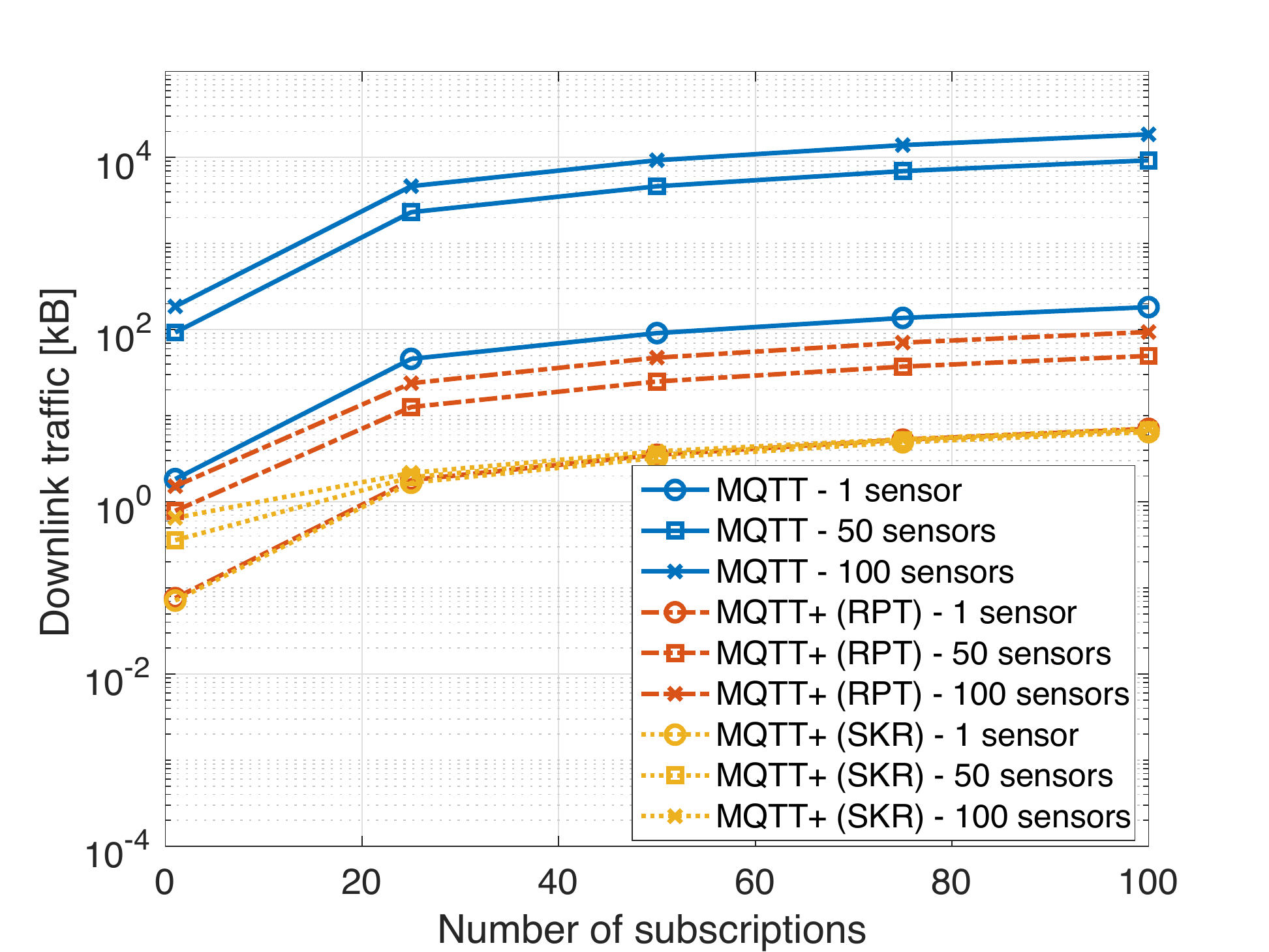}}
	\subfigure[]{\includegraphics[width=0.66\columnwidth]{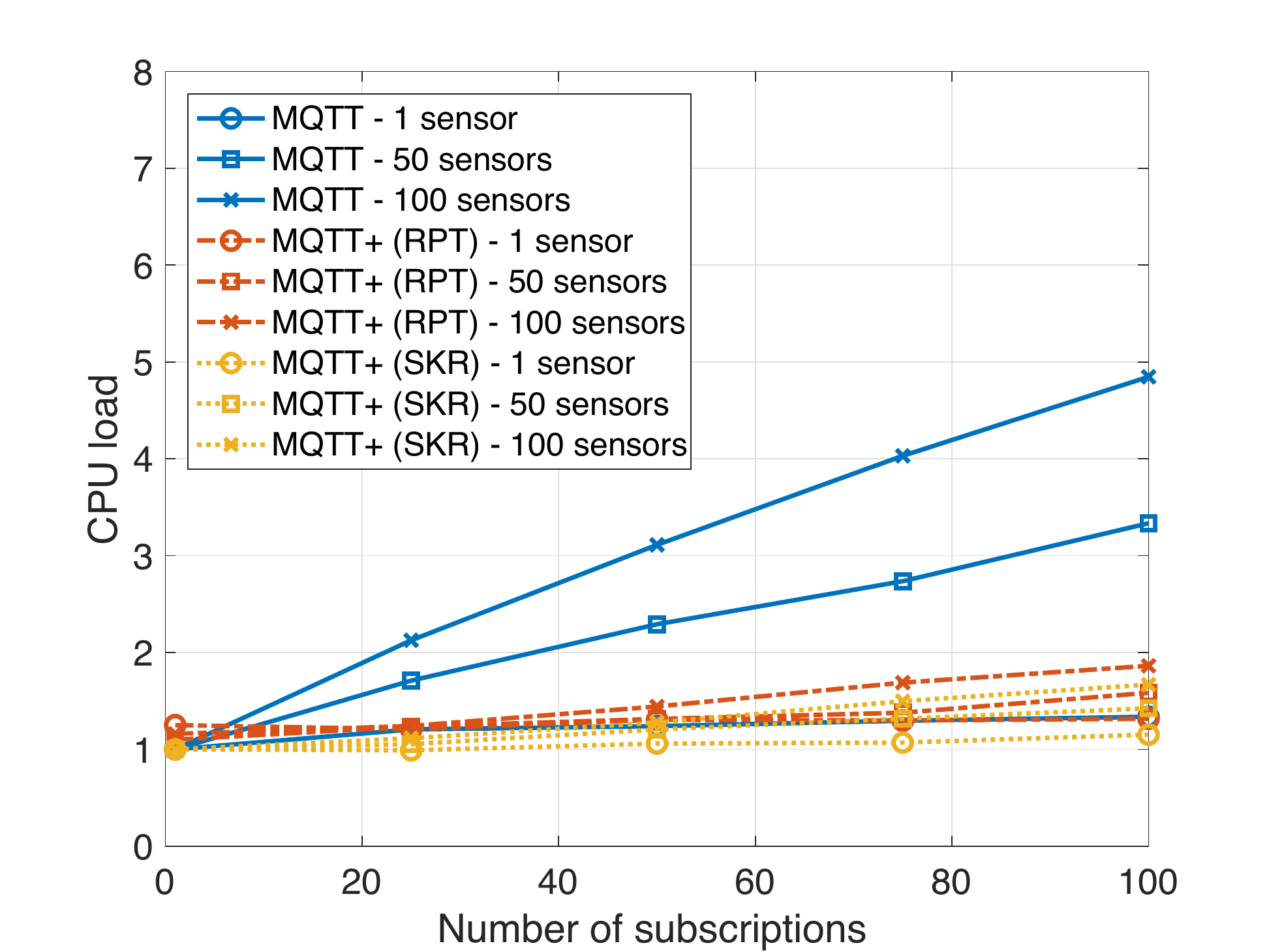}}
	\subfigure[]{\includegraphics[width=0.66\columnwidth]{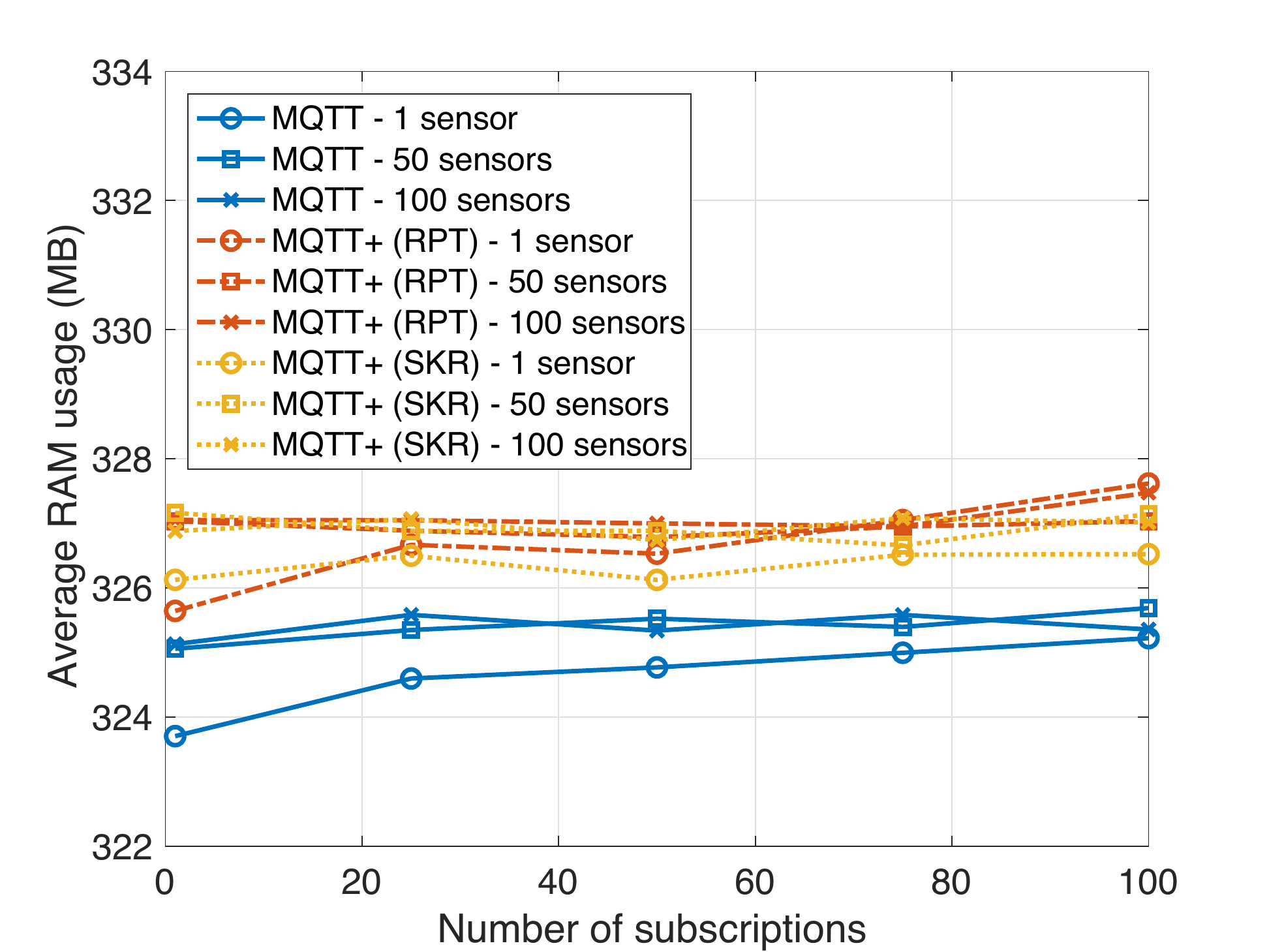}}
	\caption{MQTT vs MQTT+ spatio-temporal aggregation of scalar measurements: (a) Downlink traffic, (b) CPU Load and (c) RAM usage }
	\label{fig:numbers}
\end{figure*}

\begin{figure*}[t!]
	\centering
	\subfigure[]{\includegraphics[width=0.66\columnwidth]{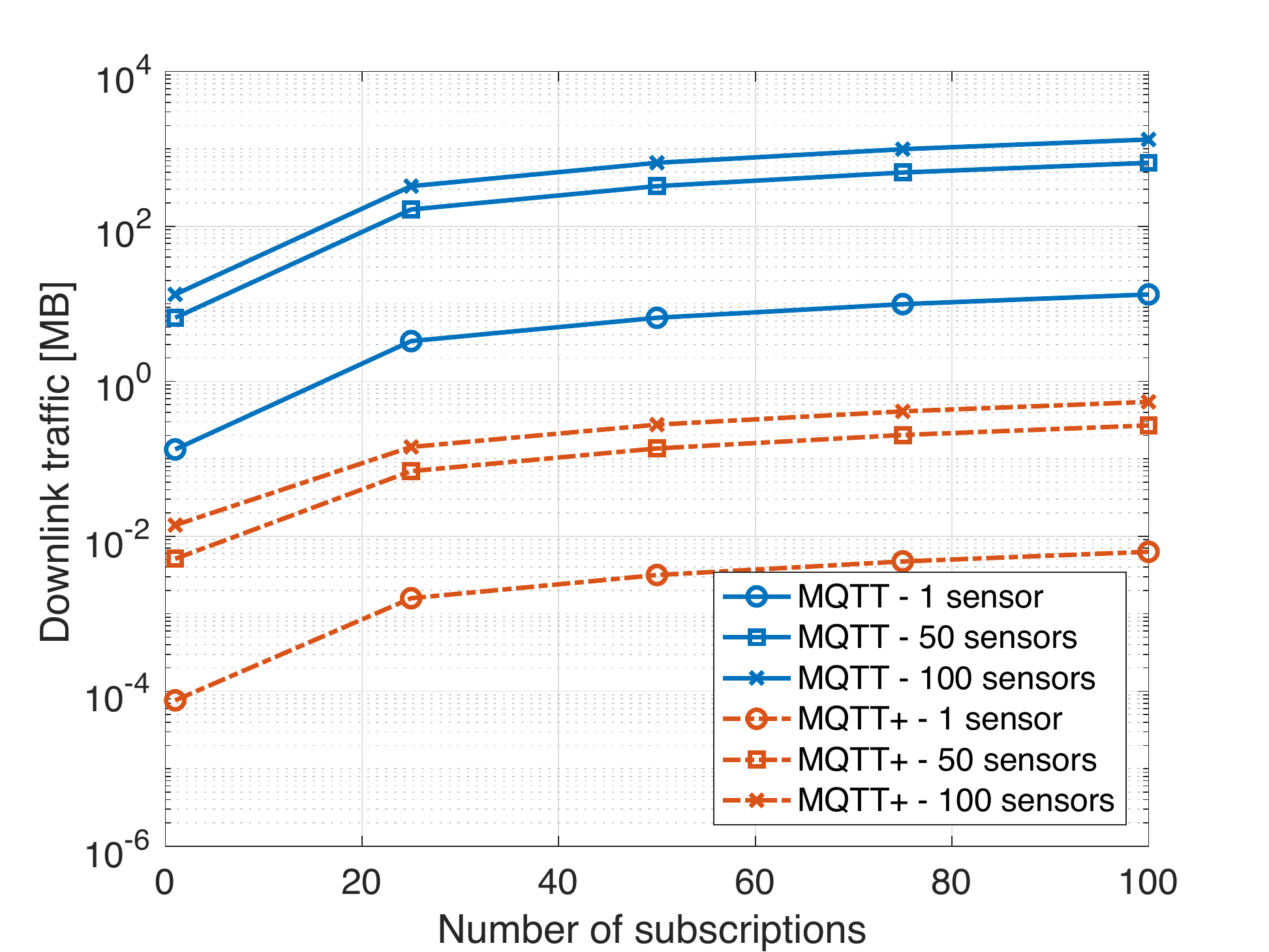}}
	\subfigure[]{\includegraphics[width=0.66\columnwidth]{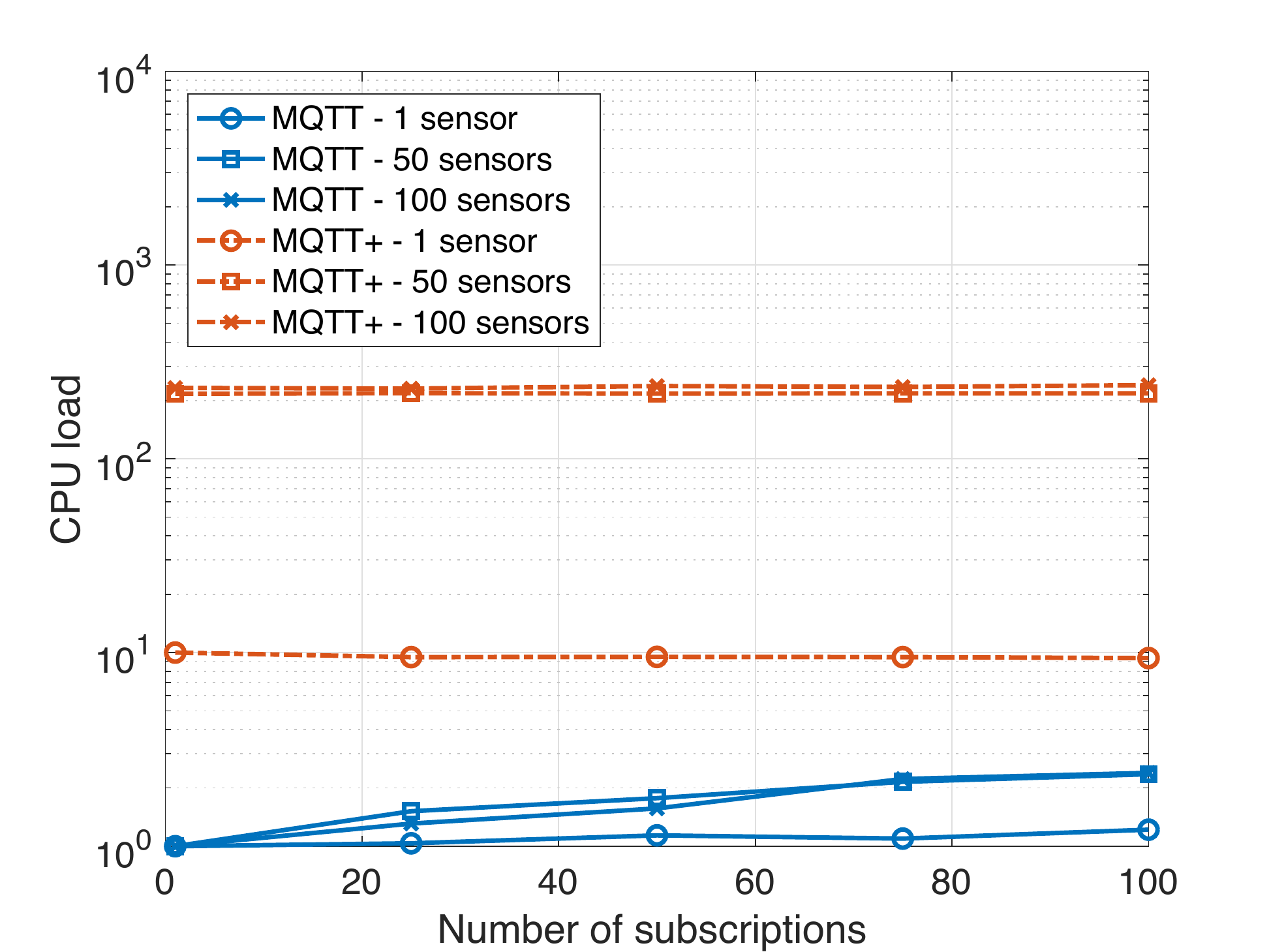}}
	\subfigure[]{\includegraphics[width=0.66\columnwidth]{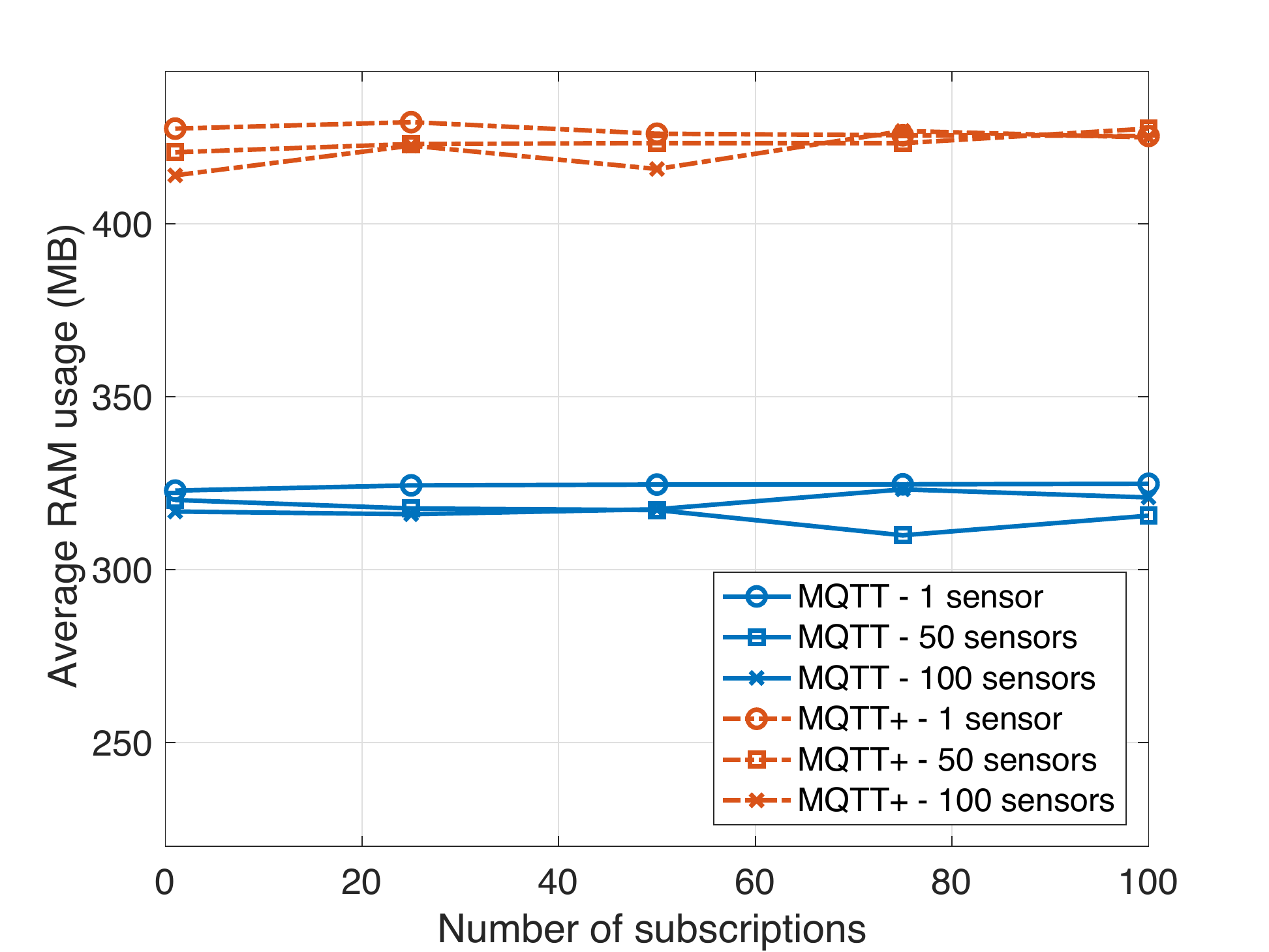}}
	\caption{MQTT vs MQTT+ processing of image data: (a) Downlink traffic, (b) CPU Load and (c) RAM usage }
	\label{fig:images}
\end{figure*}

\subsubsection{Spatio-temporal aggregation of scalar measurements}
In this scenario, the $m$ sensors publish periodically scalar data (e.g., temperature) while the $n$ subscribers are interested in a spatio-temporal aggregation of such data, rather than the individual messages. As an example, subscribers may be interested in averaging over all sensors the 15-min average of the corresponding temperature values. If the standard MQTT is used, each subscriber will receive all messages from each one of the $m$ clients, and perform the spatio-temporal aggregation onboard. Over a period of $T$ seconds, the total downlink traffic from the broker $B$ (in bytes) can be approximated as:
\begin{equation}
	B_\text{MQTT} = \lambda T (mnP)
\end{equation}
where $P$ is the length of each publish message (which includes the header $H$, the topic length $J$ and the length of the data published $K$, i.e. $P = H + J + K$ ).
Conversely, in case MQTT+ is used, the traffic forwarded by the broker can be approximated as:
\begin{equation}
	B_\text{MQTT+} = \lambda_A T (nQ)
\end{equation}
where $\lambda_A$ is the requested temporal aggregation in messages/seconds (e.g., 1 message every 900 seconds for the 15-min average), and
\begin{equation}
	Q \approx \begin{cases}
	P+O\,\, &\text{if SKR is used}\\
	P+O+Wm\,\, &\text{if RPT is used}
\end{cases}
\end{equation}
where $O$ is the length of the advanced operator used for performing aggregation. Note that in case of Replacement with Participating Topics, the downlink traffic still depends partially on the number of publishing sensors, since each individual topic level identifying each sensor (each of length $W$) must be concatenated in the payload forwarded by the broker to subscribers.

It is trivial to show that in case of spatio-temporal aggregation with Single Keyword replacement,
\begin{equation}
	\frac{B_\text{MQTT}}{B_\text{MQTT+}} \approx m\frac{\lambda}{\lambda_A}\frac{P}{P+O}
\end{equation}
i.e., the downlink traffic of an MQTT+ broker is $m\frac{\lambda}{\lambda_A}$ times lower compared to standard MQTT, with a correction factor that depends on the additional bytes used by the operator $O$ compared to the original topic $P$.

Conversely, if RPT is used, we have
\begin{equation}
	\frac{B_\text{MQTT}}{B_\text{MQTT+}} \approx m\frac{\lambda}{\lambda_A}\frac{P}{P+O+Wm}
\end{equation}
therefore the efficiency of spatio-temporal aggregation is even more decreased as the ratio between the length of the topic level used to identify sensor nodes $Wm$ and the original topic $P$ increases.

Figure \ref{fig:numbers} show the performance of MQTT and MQTT+ with SKR or RPT, measured simulating different numbers of sensors $m$ and subscribers $n$. Sensors publish numeric data ($P$ = 44) on \verb|numeric/polimi/deib/room1/sensorID|, where \verb|sensorID| is an 8 bytes identifier unique for each sensor ($W=8$). The subscribe topic is \verb|$AVG$QUARTERHOURLYAVG/numeric/polimi/deib/room1/+|,  \\($O = 21$, $\lambda_A = 1/900$ message/s) and the publishing rate of sensors is $\lambda = 1/20$ message/s.  As one can see from Figure \ref{fig:numbers}(a), spatio-temporal aggregation allows great savings in terms of the downlink network usage. The best solution is provided by MQTT+ with SKR, for which the downlink traffic from the broker is comparable with the traffic of 1 single sensor, regardless of the number of actual publishing sensors. From a computational point of view, the aggregation of numerical values does not impose a high load on the broker. Indeed, MQTT+ also allows to decrease the computational effort of a broker compared to MQTT. This is due to the fact that, in case of numerical data, the most intensive operation performed by the broker is checking the list of matching topics for each published message, in order to understand the addresses of the clients interested in data forwarding. When temporal aggregation is used, such operations are performed less frequently therefore decreasing the overall computational effort as illustrated in Figure \ref{fig:numbers}(b). In terms of memory usage, as shown in Figure \ref{fig:numbers}(c), the MQTT+ broker minimally increases the used resources, with an increment limited to 1-2\% compared to the standard MQTT broker.

\subsubsection{Processing of image data}
In the second scenario, we focus on an application in which $m$ camera sensors publish images to the broker, and the $n$ subscribers are interested in knowing the number of people present in each image. An image processing engine is therefore implemented on the MQTT+ broker and can be invoked by subscribers using the \verb|$CNTPPL| operator. Each sensor publish an image of size $I$ on \verb|image/polimi/deib/room1/sensorID| topic, and clients subscribe to \verb|$CNTPPL/image/polimi/deib/room1/+| to receive a numeric value corresponding to the number of people found. In this case the decrease in traffic of MQTT+ compared to MQTT is roughly $I/P$ with $P$ the size of a message with a numeric payload. Figure \ref{fig:images}(a) show the downlink traffic of MQTT and MQTT+ when $I/P \approx 10^2$. Observing Figure \ref{fig:images}(b) and (c), it is clear that in this case, the CPU load and memory usage on the broker is greatly affected by the intensive processing to be performed on the published images. There is therefore a tradeoff between network and computational resources, which should be carefully designed depending on the application scenarios: this may open to interesting future research directions in which the broker automatically decides whether to accept or not a data processing subscription, or rely on external resources (e.g. cloud services) to perform such processing, leading to non-trivial business model among subscribers and the owners of the broker.

\section{Related Work}\label{sec:related_work}
In the last ten years, many research studies have proposed modifications and enhancements to the MQTT protocol. One of the most popular works is the one from Hunkeler et. al which propose MQTT-SN~\cite{hunkeler2008mqtt}, a version of MQTT focused particularly on constrained wireless sensor networks. MQTT-SN do not require clients to connect to the broker through a TCP/IP connection, therefore greatly simplifying their design. Other interesting features of MQTT-SN are the possibility of using an encoded format for publishing and subscribing topics (so as to save bandwidth) and the support for clients working according to a duty cycle. Other solutions have been proposed that tackle different weaknesses of MQTT: the work in~\cite{manzoni2017proposal} tackles client mobility using memory buffers on publishers; in~\cite{singh2015secure} a lightweight encryption technique based on Elliptic Curve Cryptography is proposed to increase the security of both MQTT and MQTT-SN protocols; in \cite{govindan2015end}, authors analyse the end-to-end reliability of MQTT-SN considering several system parameters. Two very recent works show contact points with what proposed in this paper: the work in~\cite{chouali2017towards}, authors propose MQTT-CV (MQTT for communicating vehicles), in which vehicles publish sensor data and a control infrastructure is subscribed to such data. The main difference compared to MQTT is that the broker may accept some rule from the control infrastructure (e.g., forward only vehicle speed data greater or lower than a threshold). This is similar to the rule-based subscription available in the proposed MQTT+, although no details are given on how such rule-based subscriptions can be integrated in the MQTT syntax. Finally, the work in~\cite{manzoni2017proposal} proposes MQTT-NEG (Near-user Edge Gateway), a broker implementation that is able to interconnect different groups of sensors (i.e., content islands) and manage the published messages either locally (within each content island) or globally (distributing messages among different islands).

A more general body of research focuses on enhancing the capabilities of publish/subscribe systems. Li and Jacobsen propose PADRES, a pub/sub system which allows expressive and composite subscriptions tailored to the world of workflow management and business process execution. PADRES allow a subscriber to be notified when particular events (jobs in a workflow) happen in parallel, or in sequence, or repeat periodically. On the same line, Demers et al. propose Cayuga~\cite{demers2006towards} a pub/sub system allowing a user to express subscriptions spanning multiple events and supporting aggregation and parametrisation of subscriptions. The system is based on a nondeterministic finite automata and an event algebra which provides expressiveness and maps to the state of the automata. Other recent works relative to aggregation of data in generic pub/sub systems are the ones from Pandey et al. In~\cite{pandey2014distributed} and in~\cite{pandey2015minimizing} the authors propose a solution to aggregate data in a distributed way, among several brokers, together with an optimisation problem to minimise the communication cost of such distributed aggregation.
\balance

\section{Conclusion}\label{sec:conclusion}

We have proposed MQTT+, an advanced version of MQTT which allows clients to use an enhanced syntax to exploit a broker's computation power to perform different operations. MQTT+ supports rule-based subscriptions, spatio-temporal aggregation of data and advanced data processing tasks. Such basic operations can also be combined together with composite subscriptions. The MQTT+ broker is implemented starting from an existing broker implementation and tested in two different realistic scenarios, confirming the benefits of such an approach. Future research directions will further explore enhanced functionalities to be added to the broker, as well as considering the implementation of MQTT+ on top of recently proposed version of MQTT (such as MQTT-SN).

\section{Acknowledgement}
The authors would like to thank dc-square GmbH for the support received in using the HiveMQ broker.

\bibliography{sample-bibliography}

\end{document}